\documentclass[twocolumn,showpacs,preprintnumbers,amsmath,amssymb,a4paper]{revtex4}
\usepackage{textcomp}
\usepackage{latexsym}
\usepackage{amsmath}
\usepackage{amsfonts}
\usepackage{amssymb}
\usepackage{mathrsfs}
\usepackage{bm}
\usepackage{dsfont}
\usepackage[usenames]{color}
\usepackage{hyperref}
\usepackage{graphicx}

\newcommand{\stupar}{{\hat{\beta}}}

\renewcommand{\eqref}[1]{(\ref{#1})}
\newcommand{\figref}[1]{Fig.~\ref{#1}}
\newcommand{\secref}[1]{\mbox{Sec.~\ref{#1}}}

\newcommand{\abs}[1]{\left\lvert #1\right\rvert}
\renewcommand{\Re}[1]{\operatorname{Re}#1}
\renewcommand{\Im}[1]{\operatorname{Im}#1}
\newcommand{\avg}[1]{\left\langle #1 \right\rangle}

\newcommand{\nF}{N_{\text{F}}}
\newcommand{\nH}{N_{\text{H}}}
\newcommand{\project}{\Pi}
\newcommand{\integers}{{\mathbb Z}}
\newcommand{\reals}{{\mathbb R}}

\begin{document}

\title{Lyapunov stability of Vlasov Equilibria using Fourier-Hermite modes}

\author{R.~Pa\v{s}kauskas$^{1}$}
\email{rytis.paskauskas@elettra.trieste.it}
\author{G.~De~Ninno$^{1,2}$}
\affiliation{%
  $^1$Sincrotrone Trieste,
  34012 Basovizza TS, Italy \\
  $^2$University of Nova Gorica, Nova Gorica, Slovenia
}

\begin{abstract}
  We propose an efficient method to compute Lyapunov exponents and
  Lyapunov eigenvectors of long-range interacting many-particle
  systems, whose dynamics is described by the Vlasov equation. We show
  that an expansion of a distribution function using Hermite modes (in
  momentum variable) and Fourier modes (in configuration variable)
  converges fast if an appropriate scaling parameter is introduced and
  identified with the inverse of the system temperature. As a
  consequence, dynamics and linear stability properties of
  many-particle states, both in the close-to and in the far-from
  equilibrium regimes can be predicted using a small number of
  expansion coefficients. As an example of a long-range interacting
  system we investigate stability properties of stationary states of
  the Hamiltonian mean-field model.
\end{abstract}


\maketitle

\section{Introduction}\label{s:intro}
Neglecting particle correlations, the collective dynamics of many-particle systems interacting via
long-range forces obeys the Vlasov
equation~\cite{jeans1915mnras,vlasov1961,bohm1949,balescu1997}:
\begin{equation}\label{e:VLASOV}
  \dfrac{\partial f({\bf x},{\bf p},t)}{\partial t} +
        {\bf p}\dfrac{\partial f({\bf x},{\bf p},t)}{\partial {\bf x}} -
        \dfrac{\partial\varphi_f({\bf x},t)}{\partial{\bf x}}
        \dfrac{\partial f({\bf x},{\bf p},t)}{\partial {\bf p}}
        =0\,.
\end{equation}
Here $f({\bf x},{\bf p},t)$ is the single-particle density function,
while ${\bf x}$ and ${\bf p}$ are, respectively, the particle
configuration and momentum coordinates. The long-range force is
generated by the self-field
\begin{equation}\label{pote}
  \varphi_f({\bf x},t) = \int K({\bf x} - {\bf x}') \rho_f({\bf x}',t)
  d{\bf x}'\,
\end{equation}
where  $K({\bf x}-{\bf x}')$ is the particle interaction potential,
and $\rho_f({\bf x},t)=\int f({\bf x}, {\bf p},t)d{\bf p}$ is the
particle density.

There is growing evidence that the evolution of many-particle systems
with long-range interactions lends itself to interpretation in terms
of trajectories that trace a finite repertoire of collective patterns
with abrupt transitions between them, thus corroborating the claim,
stating that only a few effectively active degrees of freedom partake
in dynamics of such systems, despite the huge dimensionality of
the many-particle state space.

Presence of transitions indicates also the relevance of nonlinear
dynamical effects: instabilities, which have the capacity to strongly
alter the state of a system, even before a gradual thermalization of
the particles. Analysis of a dynamical system in terms of the
properties of its invariant structures (equilibria, manifolds,
invariant orbits) has been established as a fruitful approach to
gaining insights into the nonlinear evolution of a system~\cite{poincare}. In
particular, the stable equilibria can be identified with relevant
collective patterns, the hyperbolic manifolds of the unstable
equilibria
with pathways of transitions. With the information about the Lyapunov
spectrum and Lyapunov eigenvectors of such collective patters, control
of the systems is feasible.

Effective representation of the equilibria and of their local Lyapunov
stability properties are one of the main challenges addressed in this
Paper. Focusing on one-dimensional systems represented by density
functions with Gaussian-like tails in the momentum variable and
spatially periodic boundary conditions in the configuration variable,
we demonstrate that a Fourier-Hermite expansion of $f(x,p,t)$ can
reproduce both the Lyapunov spectrum and Lyapunov eigenvectors with a
small number of expansion coefficients. We also show that introduction
of an appropriate scaling parameter, to be identified with the inverse of
the system temperature, allows to improve the convergence of
an expansion.

As a test model, we investigate Lyapunov stability properties of
stationary states of the Hamiltonian mean-field ({HMF})
model~\cite{longrangebook2002,inagaki1993,antoni1995,yamaguchi2004,morita2006,barre2006physa,antoniazzi2007pre,latora1999,yamaguchi2007,antoniazzi2007prl1,chavanis2007epj}. 
Prior studies of the {HMF} model have revealed
non-trivial collective oscillations~\cite{morita2006} and
thermodynamical properties that pertain to a large class of long-range
interacting systems~\cite{yamaguchi2004,barre2006physa,antoniazzi2007pre,latora1999,yamaguchi2007,antoniazzi2007prl1}.

The plan of the Paper is as follows. In~\secref{s:2} we introduce the
Fourier-Hermite expansion of a density function and discuss the
 scaling parameter. The spectral equation, the
Lyapunov spectrum and Lyapunov eigenvectors are introduced
in~\secref{s:3}. Relevant facts about the {HMF} model are summarized
in~\secref{s:4}. Results, obtained from the application of our method
to the {HMF} model,
are presented in~\secref{s:5}. Finally, in~\secref{s:concludo}, we
draw conclusions.

\section{The Fourier-Hermite expansion}\label{s:2}
In the following, we focus on one dimensional systems.
Collective properties of many-particle dynamical systems are typically
more evident when the system states are represented in terms of
density functions. Further discretization of a density function
$f(x,p,t)$ in terms of a finite set of coefficients ${\bf a}=\{a_n\}$
provides an alternative representation of the system with, hopefully,
few couplings among different ${a}_n$s. The latter step can be
formalized, stating that the density function is a linear map
$\project(x,p)$ of the coefficients ${\bf a}$, given by
\begin{equation}
  f(x,p,t)=\project{(x,p)} {\bf a}(t)\,.
\end{equation}
A Galerkin projection~\cite{fornberg1995} of a density function
$f(x,p,t)$ to a set of coefficients is an example of such a map. For
the Vlasov equation, discretization in momentum variable commands
careful analysis. Therefore we first consider a partial projection
of $f(x,p,t)$:
\begin{subequations}\label{e:fexpansion}
  \begin{align}
    f(x,p,t) &= \sum_{n\geq 0} a_{n}(x,t)
    \sqrt{\stupar^{\kappa_n}} \Psi_{n}(\sqrt{\stupar} p)\,,
    \label{e:fexpansion-f} \\
    a_{n}(x,t) &= \frac{1}{\sqrt{\stupar^{\kappa_n-1}}}\int f(x,p,t)
    \Psi^n(\sqrt{\stupar} p) dp\,.
    \label{e:fexpansion-a}
  \end{align}
\end{subequations}
Here $\{\Psi_n(y)\}_{n\geq 0}$ are basis functions, and
$\{\Psi^n(y)\}_{n\geq 0}$ are basis weights. They constitute complete,
normalized, adjoint bases,
i.e. $\int\Psi^{n}(y)\Psi_m(y)dy=\delta_{nm}$. We have introduced a
free scaling function $\stupar$, which will be determined later. The
parameter $\kappa_n$ will be fixed by requiring that
$\partial\dot{a}_n/\partial a_n=0$.

In the following, we make use of the so-called asymmetrically weighted Hermite
expansion~\cite{schumer1998,armstrong1967,grant1967} (as opposed to the symmetrically
weighted Hermite expansion, discussed in~\cite{schumer1998}), defined by
\begin{subequations}
  \begin{align}
    \Psi_n(y) &= (2\pi)^{-1/2}(-d/dy)^n \exp{(-y^2/2)}\,,\\
    \Psi^n(y) &= (n!)^{-1}\exp{(y^2/2)}(-d/dy)^n
    \exp{(-y^2/2)}\,.
  \end{align}
\end{subequations}
Analysis of the Vlasov equation based on Hermite expansions in
momentum variable has been introduced
in~\cite{armstrong1967,grant1967},
with the emphasis on the accuracy of numerical solvers.

Advantages of Hermite polynomials include facts that the basis
functions decay as $\Psi_n(y)\sim\exp{(-y^2/2)}$ for large $\abs{y}$,
i.e.~in accordance with typical physical boundary conditions, and that
the recursion relations, relevant to the Vlasov equation, are simple:
\begin{subequations}\label{e:recursion}
  \begin{align}
    y\Psi_n &= A^{+}_n\Psi_{n+1} + A^{-}_n\Psi_{n-1}\,, \\
    \Psi'_{n} &=-B^{+}_n\Psi_{n+1} + B^{-}_n\Psi_{n-1}\,.
  \end{align}
\end{subequations}
In general, the density $\rho_f(x,t)$ can be expressed as
$\rho_f(x,t)={\bm\alpha}\cdot{\bf a}$, where
${\bm\alpha}=\{\alpha_n\}_{n\geq 0}$, $\alpha_n=\int \Psi_n(y)dy$.  In
the asymmetrically weighted Hermite expansion
$\rho_f(x,t)=a_0(x,t)$. Looking at~\eqref{pote}, one can see that in
this way the couplings between different momentum modes ${a}_n$ are
minimal. This fact motivates the choice of the asymmetrically weighted
expansion, whose recursion coefficients are $A^{+}_n=1$, $A^{-}_n=n$,
$B^{+}_n=1$, $B^{-}_n=0$, $\kappa_n=n+1$.

Using the expansion~\eqref{e:fexpansion}, the Vlasov
equation~\eqref{e:VLASOV} is cast as
\begin{equation}\label{e:vlasov}
  \frac{d {\bf a}}{d t} + \mathbb{V}^{-}({\bf a}){\bf a} +
  \mathbb{V}^{+}({\bf a}){\bf a} + \mathbb{S}^{-}({\bf a}){\bf a} + \mathbb{S}^{+}({\bf a}){\bf a} = 0,
\end{equation}
where $\mathbb{V}^{\pm}({\bf a})$ are defined as
\begin{multline}
  \mathbb{V}^{\pm}_{n,n'}({\bf a}) = \stupar^{\mp\frac{1}{2}}
 \delta_{n\mp 1,n'} \\
 \times
 \left(\frac{A^{\pm}_{n'}}{\sqrt{\stupar}}\frac{\partial}{\partial x} \pm
 \sqrt{\stupar} B^{\pm}_{n'}
 \frac{\partial}{\partial x}\int K(x-x') {\bm\alpha}\cdot
       {\bf a}(x',t) dx'
       \right) \,,
\end{multline}
and
\begin{multline}
  \mathbb{S}^{\pm}_{n,n'}({\bf a}) =
  (\mp) \stupar^{\mp} B_{n\mp 2}^{\pm} A_{n\mp 1}^{\pm}
  \delta_{n\mp 2,n'} \\
  \times
  \dfrac{B_1^{+} \avg{\dfrac{\partial \varphi_f}{\partial x} a_1} -
    \stupar B_3^{-}\avg{\dfrac{\partial \varphi_f}{\partial x} a_3}
  }{
    \dfrac{B_0^{+}A_{1}^{+}}{\stupar}\avg{a_0} -
    \stupar B_2^{-}A_1^{-}\avg{ a_4}
  }
\end{multline}
Here $\avg{\cdot} \equiv \int \cdot dx$.

The Vlasov dynamics typically proceeds as an interplay between two
mechanisms, the advection and the convection, and associated with them
there are two scales of spatial variations. The convection,
represented by the nonlinear term in~\eqref{e:VLASOV}, determines the
evolution of the width of the distribution function in momentum
variable. It is related to the (time-dependent) system temperature
$T(t)$:
\begin{equation}\label{e:beta}
  T(t) = \int p^2 f(x, p,t) d x d p\,.
\end{equation}
On this scale, transitions between different macroscopic states can be
observed. The advection term drives the filamentation process
independently. It is characterized by ever-thinning of inhomogeneities
(filaments) of $f$, down to the scale of fluctuations in the
underlying discrete many-particle system.

In order to represent large-scale variations of $f$ efficiently, we
focus on the former, the temperature scale. The parameter $\stupar$ in
the expansion~\eqref{e:fexpansion} can be tuned to maximize the
content of the lowest order mode $a_0(x,t)$, assumed to be
nonzero. Using \eqref{e:fexpansion}, one can show that the previous
condition is equivalent to
\begin{equation}\label{e:beta2}
  \int a_2(x,t)\,dx=0\,,
\end{equation}
which in turn corresponds to $\stupar=1/T(t)$. In this way $\stupar$
becomes a dynamical variable, whose evolution is defined by
enforcing~\eqref{e:beta2}. However, for the Lyapunov exponents of a
stationary state, derivatives of $\stupar$, and the two last terms
in~\eqref{e:vlasov}, can be neglected.

In the rest of this paper we will study the Lyapunov spectrum of smooth
distributions with periodic boundary conditions in $x$, therefore a
Fourier decomposition of $a_n(x,t)$ is adequate:
\begin{equation}\label{e:amn}
  a_{n}(x,t)=\sum_{m\in\integers} \frac{a_{mn}}{2\pi}\exp{(-\iota mx)}.
\end{equation}
With expansions~\eqref{e:fexpansion} and~\eqref{e:amn}, the system
evolution is determined by the complex coefficients
${\bf a}=\{a_{mn}\}$, where $a_{mn}=a^{\ast}_{-m,n}$.  The Vlasov matrix
$\mathbb{V}^{\pm}$ can be extended to include the expansion in $x$, by
noting that the derivative $\partial/\partial x$, acting on $a_{mn}$,
is a diagonal operator: $\partial/\partial x = -\iota m$.

For calculations, we use a finite truncation in $\nF\times \nH$
complex coefficients, and let $0 \leq m< \nF$ and $0\leq n<\nH$.

\section{Local Lyapunov Exponents}\label{s:3}
The local Lyapunov spectrum $\{\sigma_i\}$ and the corresponding Lyapunov
vectors $\{{\bm\psi}_i\}$ of a stationary state ${\bf a}_0$ are
defined by all solutions ($\sigma$,$\bm\psi$) of the spectral equation
\begin{equation}\label{e:eigen1}
  \sigma{\bm\psi} = \mathbb{A}({\bf a}_0) {\bm\psi}\,.
\end{equation}
Here $\mathbb{A}$ is the \emph{fundamental} matrix, defined by
\begin{equation}
  \mathbb{A}({\bf a}) = \iota\frac{\partial}{\partial{\bf a}}
  [\mathbb{V}^{-}({\bf a}){\bf a} + \mathbb{V}^{+}({\bf a}){\bf a}]\,.
\end{equation}

On a short time scale, the evolution of a perturbed state
$f_{\epsilon}(0)=\project {\bf a}(0)+\epsilon\project{\bm \xi}$ (where
$\epsilon$ is a small arbitrary parameter and $\bm\xi$ an arbitrary
vector) can be written as $f_{\epsilon}(t) \approx \project {\bf
  a}(0)+\epsilon\project{\bm \xi}(t)$, to the first order in
$\epsilon$. The vector $\bm\xi$ evolves according to the following
equation:
\begin{equation}
  {\bm\xi}(t) = \sum_i [\bar{\bm\psi}_i\cdot{\bm\xi}]\exp{(\iota\sigma_i t)}{\bm\psi}_i\,.
\end{equation}
Here $\{\bar{\bm\psi}_i\}$ is the adjoint basis to $\{{\bm\psi}_i\}$,
satisfying
$\bar{\bm\psi}_i{\bm\psi}_j=\delta_{ij}$. If for
some $i$, $\lambda_i \equiv \Im{(-\sigma_i)}>0$, the
corresponding eigenvector ${\bm\psi}_{i}$ is amplified exponentially
in time. It is therefore referred to as an \emph{expanding} eigenvector. The
expansion rate $\lambda_i$ is called the \emph{local Lyapunov exponent}. The
small number of local Lyapunov exponents is, as a rule, a signature of
effective low-dimensionality. The \emph{leading} local Lyapunov exponent is defined by
\begin{equation}\label{e:largest}
  \lambda = \max_i\Im{(-\sigma_i)}\,.
\end{equation}
It controls the rate of exponential divergence in time of initially
proximate states. The divergence is approximated by
\begin{equation}\label{e:expanding}
  f_{\epsilon}(x,p,t)\approx f_0(x, p ) +
  (\bar{\bm\psi}\cdot{\bm\xi}) e^{\lambda t}\project\bar{\bm\psi}\,,
\end{equation}
where $\bm\psi$ and $\bar{\bm\psi}$ are the expanding eigenvector and
its adjoint, respectively, corresponding to the spectral eigenvalue
$\sigma$ with $\Im{(-\sigma)}=\lambda$. 
If all $\sigma_i$ are real, the state is said to be \emph{spectrally}
stable.

The rate of exponential divergence between two distribution functions
is computed by monitoring the evolution of their difference. This
requires the introduction of the concept of the distance between two
distribution functions. We define the distance $d(f_\epsilon,f_0)$ as
\begin{equation}\label{e:distance}
  d(f_\epsilon(t),f_0)\\
  = \left(\sum_{mn}\abs{(a_{\epsilon})_{mn}-(a_{0})_{mn}}^2\right)^{1/2}.
\end{equation}

\section{The {HMF} model}\label{s:4}
\subsection{Equations of Motion}
The Hamiltonian mean-field ({HMF})
model~\cite{longrangebook2002,inagaki1993,antoni1995} is a continuous
time model of a one-dimensional ``gas'' of $N$ globally coupled
particles on a circle, with coordinates
$-\pi\leq x\leq\pi$, momenta $p\in\reals$ and the ``magnetization''
${\bf m}(x)=(\cos{x},\sin{x})$.  The mean
magnetization is defined by $\avg{{\bf m}}=N^{-1}\sum_i{\bf m}_i$, where ${\bf
  m}_i={\bf m}(x_i)$. The {HMF} Hamiltonian is
\begin{equation}\label{e:hmf-H}
  H( \{ x_i,p_i\}_{i=1}^{N})
  = \sum_{i=1}^{N}\dfrac{p_i^2}{2} +\dfrac{1}{2N}\sum_{i,j=1}^{N}
  \left[ 1 - {\bf m}_i\cdot{\bf m}_j\right]\,.
\end{equation}
It determines the particle evolution by
\begin{equation}\label{e:hmf-EOM}
  \frac{d^2 x_i}{dt^2} + \avg{\bf m}\wedge {\bf m}_i = 0\,.
\end{equation}
\begin{figure}
  \begin{center}
    \includegraphics{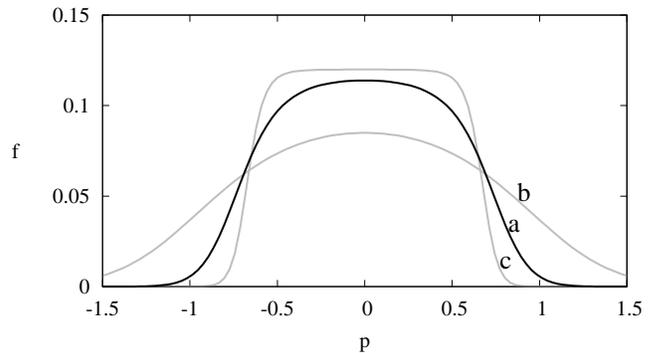}
    \caption{\label{f:1}
      Several homogeneous stationary states $f_{0}(p;\alpha,\beta)$,
      corresponding to fixed $w_p=1.3262$. Energies are $U_a=0.6$,
      $U_b=0.72$, $U_c=0.577$ (see also~\figref{f:4}.)}
  \end{center}
\end{figure}
In the limit $N\rightarrow\infty$, fixed energy per particle
$U=H/N$,
\begin{equation}
  U=\frac12\avg{p^2} + \frac{1-\abs{\avg{\bf m}}^2}{2}\,,
\end{equation}
correlations between particles yield to collective phenomena. In
this, collisionless limit, the Vlasov equation for the {HMF} assumes
the form
\begin{equation}\label{e:hmf-vlasov}
  \frac{\partial f}{\partial t} + p\frac{\partial f}{\partial x}
   - {\bf m}_f(t)\wedge{\bf m}(x)
   \frac{\partial f}{\partial p} = 0\,,
\end{equation}
where $f(x,p,t)$ is the single-particle distribution function. 
The mean-field magnetization is defined by 
 ${\bf m}_f(t)=\int {\bf m}(x) f(x,p,t) dx dp$.

\subsection{Homogeneous Stationary States}
We consider a two-parameter family of homogeneous (${\bf m}_f=0$)
distributions $f_0(p;\alpha,\beta)$, given by
\begin{equation}\label{e:lbm0}
  f_\text{0}(p;\alpha,\beta) =
  \dfrac{1}{2\pi w_p}\dfrac{1}{1+\exp{(\beta p^2/2 - \alpha)}}\,
\end{equation}
This distribution is a special case of a class of
distributions, defined by $f\sim [1+\exp{(\beta H-\alpha)}]^{-1}$,
shown to be important in the statistical treatment of the ``violent
relaxation'' processes~\cite{lynden-bell1967}. Their relevance to the
{HMF} model has been discussed in~\cite{antoniazzi2007pre}.

Assuming that $f_0$ is normalized, $\int f_{0}(p)\,dp=2\pi$, the
parameters $\alpha$, $\beta$ are related to $w_p$, $U$ by the
self-consistency conditions~\cite{chavanis2007epj}: $\beta w_p^2 =
2\pi [F_{-1/2}(\alpha)]^2$, $U = F_{1/2}(\alpha)/2\beta
F_{-1/2}(\alpha) + 1/2$. Three examples of $f_0(p; \alpha,\beta)$ with
different values of $U$ are shown in~\figref{f:1}.

The effective temperature $\stupar$,
determined by~\eqref{e:beta2}, is
\begin{equation}\label{e:stupar}
  \stupar=\frac{\beta F_{-1/2}(\alpha)}{F_{1/2}(\alpha)}\,,
\end{equation}
where
\begin{equation*}
  F_j(\alpha) = \dfrac{1}{\Gamma(j+1)}\int_{0}^{\infty}
  \dfrac{t^j}{\exp{(t-\alpha)}+1}\,
\end{equation*}
are the standard Fermi-Dirac integrals.

\begin{figure}
  \begin{center}
    \includegraphics{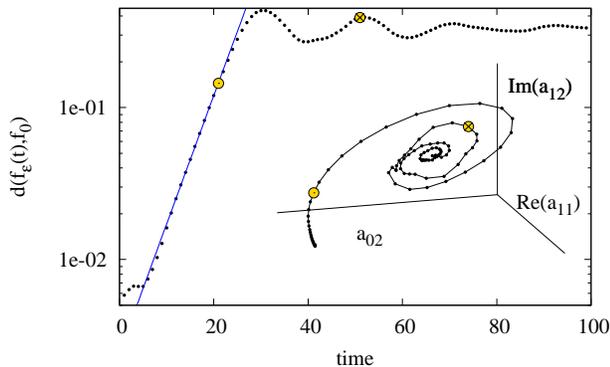}
    \caption{\label{f:2} Temporal evolution of a perturbation of the
      stationary state (a) of~\figref{f:1}, represented by the
      distance~\eqref{e:distance}, and as a trajectory in the
      coefficient space in the inset.  The thick open circle marks the
      state at $t=20$, while the thick crossed circle marks the state
      at $t=50$.  A fit of the exponential regime predicts
      $\lambda=0.1947\pm 0.0013$.  The inset shows a projection of the
      trajectory in ($a_{0,2},\Re{(a_{1,1})},\Im{(a_{1,2})}$).  The
      initial perturbation of the stationary state is obtained by
      Monte Carlo integration of the distribution function
      $f_0(p)$~\eqref{e:lbm0} with one million particles, which is further used
      as an initial condition to evolve~\eqref{e:hmf-EOM}.  }
  \end{center}
\end{figure}

In Ref.~\cite{chavanis2007epj}, it has been demonstrated that, in the
parameter plane ($w_p$, $U$), the stable and the unstable stationary
states~\eqref{e:lbm0} are separated by the boundary curve
$\ell_{\text{c}}(\alpha)=(w_{p,c}(\alpha), U_{\text{c}}(\alpha))$.
Defining $G_{j}(x,\alpha)$ for $\alpha\in\reals$ and
$x\in\reals^{+}$ by
\begin{equation}\label{e:Gdef}
  G_j(x,\alpha) = \frac{1}{\Gamma(j+1)}\int_{0}^{\infty}
  \frac{\exp{(t-\alpha)}}
       {[1 + \exp{(t-\alpha)}]^2}\frac{ t^{j+1} dt}{t+x}\,,
\end{equation}
we express $\ell_{\text{c}}$ as
\begin{multline}\label{e:boundary}
  \ell_{\text{c}}(\alpha) = \left(
  \left[\pi F_{-1/2}(\alpha) G_{-1/2}(0,\alpha) \right]^{1/2},\right.\\
  \left.\frac{F_{1/2}(\alpha) G_{-1/2}(0,\alpha)}
       {4[F_{-1/2}(\alpha)]^2} + \frac12\right)\,.
\end{multline}
In Ref.~\cite{chavanis2007epj} it has also been demonstrated that for
a fixed value of $w_p$, $U\geq U_{\text{min}}(w_p)=1/2 +
w_p^2/24$. This defines the limiting boundary curve
$\ell_{\text{min}}(w_p)=(w_p, w_p^2/24+1/2)$.  Note that approaching
$\ell_{\text{min}}$ corresponds to taking the asymptotic limit of
$\alpha\rightarrow\infty$ of the distribution
function~\eqref{e:lbm0}. In this limit $f_0$ tends to the
``water-bag'' distribution, defined by $f_{\text{wb}}(p) = (2\pi
w_p)^{-1} \Theta(\abs{p}-w_p/2)$, where $\Theta(p)$ is the Heaviside
function.  Close to this boundary, differentiability properties of
$f_0$ deteriorate, and the convergence of the coefficient expansion is
slower.

\begin{figure}
  \begin{center}
    \includegraphics[width=8.5cm]{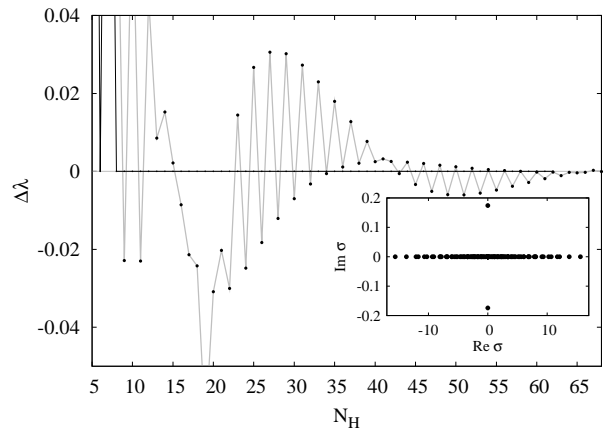}
    \caption{\label{f:3}
      The difference $\Delta \lambda$ between the largest Lyapunov
      exponent~\eqref{e:largest} and the exact value~\eqref{e:lmap}, as a
      function of the truncation order $\nH$ in Hermite polynomials. 
      Here $\nF=5$. Two cases are shown: the unstable state (a) (dotted line) and
      the stable state (b) (continuous line). The inset shows the complete
      Lyapunov spectrum of the unstable state (a).
    }
  \end{center}
\end{figure}

\begin{figure*}
  \begin{center}
    \includegraphics[width=\textwidth]{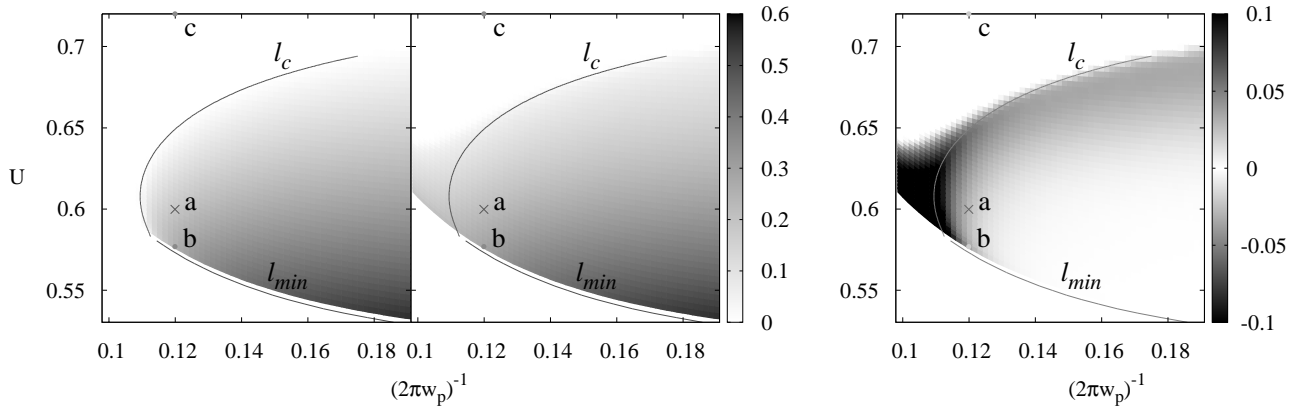}
    \caption{\label{f:4} Comparison between the exact Lyapunov map and
      computation using the truncated spectral
      equation. The left panel shows the Lyapunov
      map~\eqref{e:lmap}. The middle panel shows the results, obtained
      from the spectral equation~\eqref{e:eigen2} with $\nF=5$, and averaging over $\nH$
      between 31 and 61. Their difference is shown in the right
      panel. Special values of parameters, discussed
      in this paper, are indicated by (a), (b), (c).  }
  \end{center}
\end{figure*}

In the following we will compare our calculations of Lyapunov exponents with
results on phase transitions for the {HMF} model, discussed in
Ref.~\cite{antoniazzi2007pre,chavanis2007epj}. 

Figure~\ref{f:2} shows the divergence from an equilibrium of a perturbed
homogeneous stationary state, corresponding to the curve (a)
in~\figref{f:1}. The evolution has been obtained by integrating~\eqref{e:hmf-EOM}. The initial exponential divergence
is clearly displayed, together with the subsequent saturation to a
quasi stationary state, characterized by low-frequency
oscillations. The inset of~\figref{f:2} shows the system dynamics,
projected in coefficients
($a_{0,2}$,$\Re{(a_{1,1})}$,$\Im{(a_{1,2})}$). As it can be seen, the
stretch of the (almost linear) initial trajectory, slightly bending
towards the attractor just before the reference point at $t=20$, shows
the qualitative features of the long-term dynamics.

\section{Results and Discussion}\label{s:5}

Linearization of the Vlasov Equation~\eqref{e:VLASOV} around the
family of distributions~\eqref{e:lbm0} shows that the Lyapunov
spectrum comprises a continuum of real eigenvalues, associated
with spectrally stable modes, as well as a finite number of imaginary
eigenvalues, to be associated with unstable collective modes. In the
case of homogeneous distributions, Lyapunov exponents of the
collective modes can be expressed succinctly, using the plasma
dispersion function $\varepsilon(\sigma)$, as an equation
$\varepsilon(\sigma)=0$~\cite{bohm1949,balescu1997}. The 
plasma dispersion function for the {HMF} model has been derived
in~\cite{antoni1995}:
\begin{equation}\label{e:dispersion}
  \varepsilon(\sigma) =1+
  \pi\int_{-\infty}^{\infty}\dfrac{1}{p+\sigma}
  \dfrac{\partial f_0(p)}{\partial p} dp,
\end{equation}    
where $f_0$ is given by~\eqref{e:lbm0}. 
The boundary between stable and unstable stationary states is determined
by $\varepsilon(0)=0$. The result is the
curve $\ell_{\text{c}}$, given by~\eqref{e:boundary}.

The local (real and positive) Lyapunov exponent $\lambda$ is found as
a solution of $\varepsilon(\iota\lambda) = 0$.
We define this solution as the Lyapunov map:
\begin{equation}\label{e:lmap}
  \lambda (w_p,U) = \sqrt{\frac{2}{\beta} G_{-\frac12}^{-1} \left( \dfrac{2 F_{-\frac12}(\alpha)}{\beta},\alpha\right)}\,,
\end{equation}
where $G^{-1}_{j}(x,\alpha)$ is the inverse of $G_{j}(x,\alpha)$
(given by~\eqref{e:Gdef}) with respect to the variable $x$, while 
$\alpha=\alpha(w_p,U)$ and $\beta=\beta(w_p,U)$ are expressed in terms
of $w_p$ and $U$ using the self-consistency conditions. 

For non-homogeneous stationary distributions, the dispersion
relation does not have a closed form, such
as~\eqref{e:dispersion}, and solving~\eqref{e:eigen1}
is the only direct way to obtain the Lyapunov spectrum.
Equation~\eqref{e:eigen1} specialized for the {HMF} model reads
\begin{multline}\label{e:eigen2}
  \sigma{\psi}_{m,n} =
        {m}\left( \frac{{\psi}_{{m},{n-1}}}{\stupar} + ({n}+1)
        {\psi}_{m,n+1}\right) \\
        + \sum_{k=\pm 1} \frac{k}{2}
        [a_{k,0} {\psi}_{ m - k, n-1} + a_{m-k,n-1}{\psi}_{k,0} ]\,.
\end{multline}

\begin{figure*}
  \begin{center}
    \includegraphics[width=\textwidth]{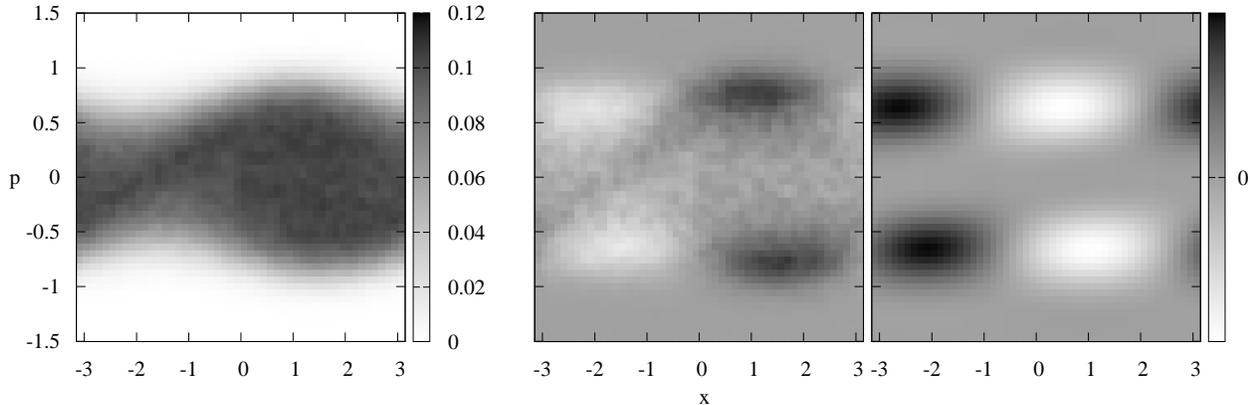}
    \caption{\label{f:5}%
      Exponential amplification of a perturbed stationary state (a). 
      The left panel shows the distribution $f(t)$ at $t=20$
      (indicated by a thick dot in~\figref{f:2}.)
      In order to verify~\eqref{e:expanding}, 
      the center panel shows the difference
      between $f(20)$ and the initial distribution.
      The right panel shows the expanding eigenvector,
      corresponding to $\lambda_a$, computed
      using~\eqref{e:eigen2} with $\nF=5$, $\nH=11$. The phase
      difference in the configuration
      coordinate $x$ between the right and middle panels is due to the
      translation invariance of the {HMF} model.}
  \end{center}
\end{figure*}
Figure~\ref{f:3} shows the difference between the Lyapunov
exponent~\eqref{e:largest}, and the exact value, given by~\eqref{e:lmap}, as a
function of the truncation order $\nH$ in Hermite polynomials. The
number of Fourier modes is fixed at $\nF=5$. Two cases are shown,
corresponding to the unstable state (a) (exact value
$\lambda_{\text{a}}=0.189549$) and to the stable state (b)
($\lambda_{\text{b}}=0$), in~\figref{f:1}. As it can be seen, $\nH\geq
7$ allows to predict real Lyapunov spectrum of the stationary state
(b), and to predict a positive Lyapunov exponent of the unstable
stationary state (a) qualitatively. Given that the sources of errors
include both the finite truncation error, and errors in the
approximation of the coefficients (computed using Monte-Carlo
integration), the convergence to the exact value $\lambda_a$ can be
considered as satisfactory. The inset of~\figref{f:3} shows the
complete Lyapunov spectrum of the unstable state (a), including a
single local Lyapunov exponent. All remaining eigenvalues have a
vanishing imaginary part and tend to cover the real axis densely.

In~\figref{f:4} we display the Lyapunov map $\lambda(w_p,U)$. The
exact map~\eqref{e:lmap}, the approximation computed
via~\eqref{e:eigen2}, and their difference are shown in the left,
center and right panels, respectively. The center panel was computed
by averaging the largest Lyapunov exponent over the order of
truncation, $\nH$, between $\nH=31$ and $\nH=61$, and with the fixed
$\nF=5$.  The agreement is fully satisfactory for the high-energy
stable stationary states. Indeed, consistently with what is reported
in~\figref{f:3}, we have verified that for $\nH \geq 7$ the computed
by~\eqref{e:eigen2} Lyapunov spectrum in this region is always real,
and therefore predictions of spectral stability are
unambiguous. Similarly, apart from the region close to the
intersection between $\ell_{\text{min}}$ and $\ell_{\text{c}}$, one
can see that also the unstable states are predicted unambiguously. The
largest Lyapunov exponent converges to the exact value as the
truncation size is increased.  As for the error in the intersection
region, analysis, similar to the one shown in ~\figref{f:3},
demonstrates that close to $\ell_{\text{c}}$ the prediction of
$\lambda$ is more sensitive to errors in the coefficients
$(a_0)_{mn}$. Moreover, deterioration of results computed with a fixed
truncation order $\nH$ close to $\ell_{\text{min}}$ can be anticipated
because of differentiability properties of $f_0$.

\begin{figure*}
  \begin{center}
    \includegraphics{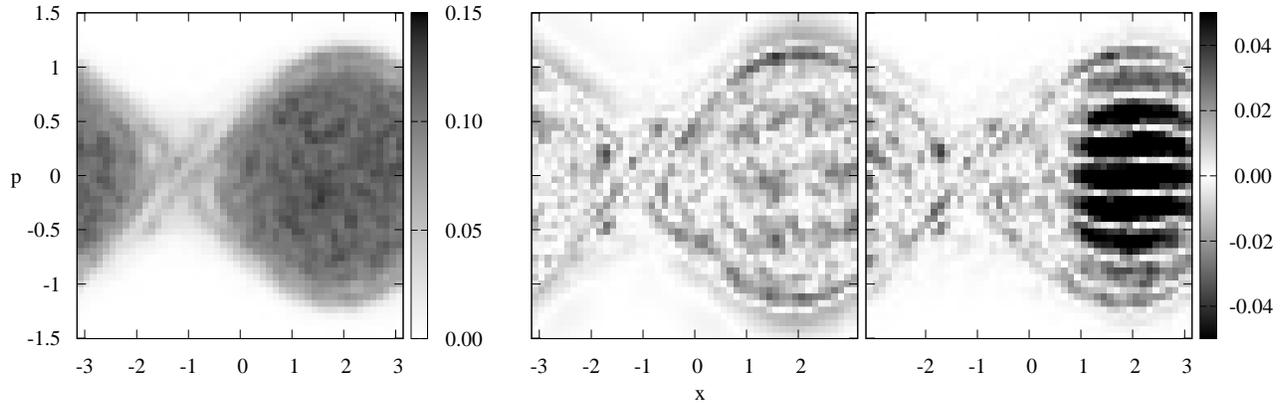}
    \caption{\label{f:6}
      The left panel shows the reference distribution $f(t)$ at $t=50$, at which
      time the effective temperature~\eqref{e:beta} is $T\approx
      1/3.5$. The center and the right 
      panels show the difference between the reference distribution
      and the Fourier-Hermite expansion~\eqref{e:fexpansion}, with
      $\stupar=3.5$ and $\stupar=10$, respectively. In both cases $\nF=5$, $\nH=11$.
      The point of reference $t=50$ is also indicated by a thick crossed circle in~\figref{f:2}.
}
  \end{center}
\end{figure*}

For comparison, the exact Lyapunov exponent for the state (a)
$\lambda_{\text{a}}=0.1895$. Simulation of one million particles
(see~\figref{f:2}) gives $\lambda=0.1947$. The convergence to
$\lambda_{\text{a}}$ of the calculation based on Fourier-Hermite
expansion is displayed in~\figref{f:3}. On can see that our method is
in good agreement with both the $N$-particle simulation and the exact
calculation.

The full solution of~\eqref{e:eigen1} provides detailed information
about local directionality of transitions associated to each
eigenvalue. In the presence of a single positive local Lyapunov
exponent, or if one local Lyapunov exponent is larger than the
remaining ones, this directionality is expressed
by~\eqref{e:expanding}. To test accuracy of our method with respect to
calculation of expanding eigenvectors, in Fig.~\ref{f:5} we explore
the temporal divergence of a perturbed stationary state (a). The
distribution function, obtained by directly
integrating~\eqref{e:hmf-EOM} up to $t=20$, to near the breakdown of
the regime of exponential divergence (see~\figref{f:2}), is displayed
in the left panel of~\figref{f:6}. Subtraction 
of the initial state from the latter is shown in the central panel. A
direct comparison with the unstable expanding mode, calculated using
truncated~\eqref{e:eigen2} with $\nF=11$, $\nH=5$ (see the right
panel), demonstrates that also Lyapunov eigenvectors are predicted
accurately.

As a final step, we tested the ability of the method to predict the
distribution function after the exponential regime, i.e. outside the
limit of validity of~\eqref{e:eigen2}. Results are summarized
in~\figref{f:6}. Considering again the perturbed stationary state (a),
the distribution function at $t=50$ is shown in the left panel of~\figref{f:6}
(indicated by a thick crossed circle in~\figref{f:2}.) The effective temperature
of the distribution is $T\approx 1/3.5$. The difference between a
representation using fixed low-order truncation ($\nF=5$, $\nH=11$) is shown in
the center panel with $\stupar=3.5$, and in the right panel with $\stupar=10$.
As it can be seen, the case in which $\stupar$ is tuned at the inverse of the system
temperature results in smaller errors of the approximation. The
agreement using a detuned $\stupar$ can be improved by increasing the
order of the Hermite expansion.

\section{Conclusions}\label{s:concludo}
We have proposed an efficient method to compute Lyapunov exponents and
Lyapunov eigenvectors of collective states in long-range interacting
many-particle systems, whose dynamics is described by the Vlasov
equation. The method is based on expanding the distribution function
in a Fourier-Hermite basis, with a scaled momentum variable. Having
tuned the scaling parameter to maximize the content of the lowest
order coefficient of the expansion, we have demonstrated that the
Lyapunov exponent and Lyapunov eigenvectors converge fast to the
values predicted by direct many-particle simulations and by the exact
analytical formula. In the last part of the paper, the ability of the
method to predict the evolution of the distribution function in
the far-from-equilibrium regime has been demonstrated.  As an example
of a long-range interacting system, the Hamiltonian mean-field model
has been considered. In this context, stability properties of a
two-parameter family of homogeneous distributions over a wide range of
dynamical conditions has been investigated. Our conclusion is that
linear stability properties and collective aspects of dynamics of the {HMF} model can be
represented and computed using a small number of modes, 
and thus a small numerical effort.

\newcommand{\noopsort}[1]{} \newcommand{\printfirst}[2]{#1}
  \newcommand{\singleletter}[1]{#1} \newcommand{\switchargs}[2]{#2#1}

\end{document}